\documentclass[12pt,a4paper]{article}
\usepackage{graphicx}
\usepackage{times}
\title{\bf The interactions of winds from massive young stellar objects}
\author{E.~R.~Parkin$^{1,2}$, J.~M.~Pittard$^2$, M.~G.~Hoare$^2$, N.~J.~Wright$^{3}$, J.~J.~Drake$^{3}$\\
\vspace{1cm}\\
\normalsize $^1$ Institut d'Astrophysique et de G\'{e}ophysique, Universit\'{e} de Li\`{e}ge, Belgium\\ 
\normalsize $^2$ School of Physics and Astronomy, The University of Leeds, UK\\
\normalsize $^3$ Harvard-Smithsonian Center for Astrophysics, USA}
\date{\mbox{}}
\begin{document}
\maketitle
\pagestyle{empty}
%
%
\def\bull{\vrule height .9ex width .8ex depth -.1ex}
\makeatletter
\def\ps@plain{\let\@mkboth\gobbletwo
\def\@oddhead{}\def\@oddfoot{\hfil\tiny\bull\quad
``The multi-wavelength view of hot, massive stars''; 39$^{\rm th}$ Li\`ege Int.\ Astroph.\ Coll., 12-16 July 2010 \quad\bull}%
\def\@evenhead{}\let\@evenfoot\@oddfoot}
\makeatother
%
%
\def\beginrefer{\section*{References}%
\begin{quotation}\mbox{}\par}
\def\refer#1\par{{\setlength{\parindent}{-\leftmargin}\indent#1\par}}
\def\endrefer{\end{quotation}}
%
%
{\noindent\small{\bf Abstract:} The supersonic stellar and disk winds
  possessed by massive young stellar objects will produce shocks when
  they collide against the interior of a pre-existing bipolar cavity
  (resulting from an earlier phase of jet activity). The shock heated
  gas emits thermal X-rays which may be observable by spaceborne
  observatories such as the Chandra X-ray Observatory. Hydrodynamical
  models are used to explore the wind-cavity interaction. Radiative
  transfer calculations are performed on the simulation output to
  produce synthetic X-ray observations, allowing constraints to be
  placed on model parameters through comparisons with
  observations. The model reveals an intricate interplay between the
  inflowing and outflowing material and is successful in reproducing
  the observed X-ray count rates from massive young stellar objects.
}
%
%
\section{Introduction}

Observational and theoretical advances have provided increasing
evidence that massive star formation is not merely a scaled-up version
of lower-mass star formation. However, there are some
similarities. For instance, both involve outflows (Garay \& Lizano
1999; Reipurth \& Bally 2001; Banerjee \& Pudritz 2006, 2007) and
bipolar cavities are commonly observed around both high and low mass
young stellar objects (Garay \& Lizano 1999). The widths of IR
recombination line emission observed from massive young stellar
objects (MYSOs) indicates the presence of dense ionized outflows with
velocities ranging from 100 to $>340\;{\rm km\thinspace s}^{-1}$
(Drew, Bunn, \& Hoare 1993; Bunn, Hoare, \& Drew 1995). Further
confirmation of outflows has come from high angular resolution radio
observations (e.g. Hoare et al. 1994; Hoare 2006; Curiel et
al. 2006). One explanation would be a disk wind driven by radiation
pressure at the surface of the disk (e.g. Drew et al. 1998).

X-rays have been detected from deeply embedded MYSOs in star forming
regions by the \textit{Chandra X-ray observatory} (hereafter
\textit{Chandra})(e.g. Broos et al. 2007; Wang et al. 2007). Preibisch
et al. (2002) found that the X-ray emission from IRS3 A and C, with a
count rate of $0.166\pm 0.041\;{\rm ks^{-1}}$ for the former, could
not be explained by the standard scenario for massive stars (i.e. wind
embedded shocks produced by instabilities inherent in
radiatively-driven winds - see Owocki, Castor, \& Rybicki 1988), yet
the estimated stellar masses of these objects implies they will have
radiative outer envelopes which poses problems for the generation of
X-rays through magnetic star/disk interactions.

A potential source of X-rays which has not been considered before is
the collision between the stellar and disk winds and the infalling
envelope. We explore this scenerio using hydrodynamical models and
find that the interaction of the stellar and disk wind with the cavity
wall produces shock heated plasma which emits X-rays in agreement with
Chandra observations (Parkin et al. 2009).

The remainder of this work is structured as follows: in
\S~\ref{sec:model} we describe the winds-cavity model, in
\S~\ref{sec:results} we present the result of the hydrodynamic
simulations and X-ray calculations, and in \S~\ref{sec:conclusions} we
close with our conclusions.

\begin{figure}[h]
\centering
\includegraphics[width=12cm]{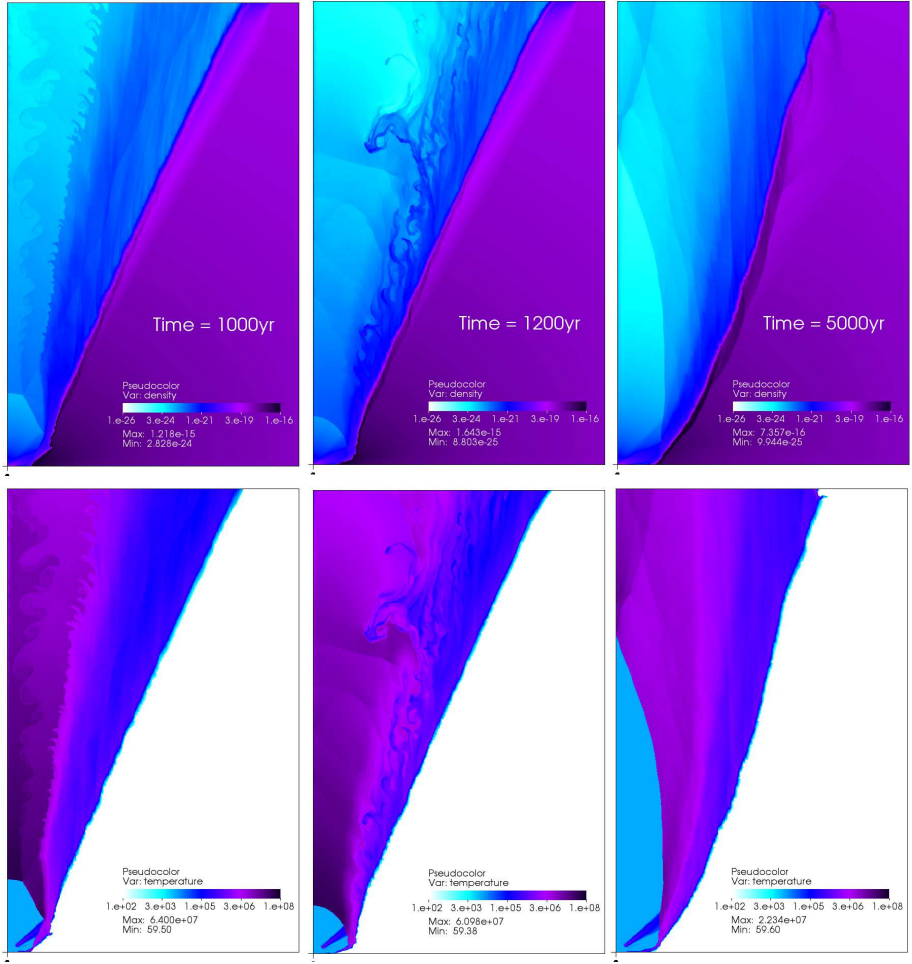}
\caption{Density (top row) and temperature snapshots (bottom row) from
  our fiducial model at a simulation time of $t=1000\;$(left), 1200
  (middle), and 5000 yr (right). The grids extend to $x=5 \times
  10^{16}\;$cm and $y= 8 \times 10^{16}\;$cm. The central star is
  located in the bottom left corner. The reverse shock can be seen in
  the lower left of the grid, where the enclosed preshock wind has a
  temperature of $T=10^{4}\;$K. \label{fig:snapshots}}
\end{figure}

\section{The winds-cavity interaction model}
\label{sec:model}

The model consists of a MYSO situated at the centre of a previously
evacuated bipolar cavity which is surrounded by infalling molecular
cloud material. We include the stellar wind and a disk wind which
emanates from the surface of the circumstellar accretion disk; both
are assumed to be at terminal velocity. Due to the spatial scales
under consideration we do not attempt to model the radiative driving
of the winds as this requires high spatial resolution in the vicinity
of the star/disk (e.g. Proga et al. 1998). For simplicity we adopt an
angle dependent wind prescription based on the models of Proga et
al. (1998), Drew et al. (1998), and Sim et al. (2005), whereby the
stellar wind occupies a region from a polar angle $0-60^{\circ}$ and
the disk wind occupies the region from $60^{\circ}-85^{\circ}$. The
density and velocity distibutions for the infalling cloud material are
described by the equations of Terebey et al. (1984), and a simple
analytical prescription similar to that of Alvarez et al. (2004) is
used to determine the morphology of the pre-existing outflow
cavity. We model the winds-cavity interaction in 2D cylindrical
symmetric coordinate system using {\sc VH-1} (Blondin et
al. 1990). The code uses the piecewise-parabolic method of Colella \&
Woodward (1984) to solve the hydrodynamic equations on a fixed
grid. Further details of the model can be found in Parkin et
al. (2009).

We consider a 30$\;{\rm M_{\odot}}$~O8V star with a mass-loss rate of
$10^{-7}\;{\rm M_{\odot}~yr^{-1}}$ and terminal wind speed of
2000$\;{\rm km~s^{-1}}$. The disk wind has a mass-loss rate of
$10^{-6}\;{\rm M_{\odot}~yr^{-1}}$ and a terminal wind speed of
400$\;{\rm km~s^{-1}}$. The unshocked winds are assumed to be at a
temperature of $10^{4}\;$K. The mass infall rate for the cloud is
$2\times 10^{-4}\;{\rm M_{\odot}~yr^{-1}}$ and the cavity opening
angle is 30$^{\circ}$ . The winds-cavity interaction was followed for
a simulation time of $t = 5000\;$yrs.

To allow a comparison to be made between \textit{Chandra} X-ray
observations of MYSOs and the simulations we calculate attenuated
X-ray fluxes. The emissivity values used are for optically thin gas in
collisional ionization equilibrium obtained from look-up tables
calculated from the \textsc{MEKAL} plasma code (Liedahl, Osterheld, \&
Goldstein 1995 and references there-in). Ray-tracing calculations are
performed with an inclination angle (to the pole) of $60^{\circ}$.

In this work we only consider one set of parameters, and we refer the
reader to Parkin et al. (2009) for a detailed parameter space
exploration.

\section{Results}
\label{sec:results}

Fig.~\ref{fig:snapshots} shows the spatial distribution of material in
the simulation; the disk wind lines the cavity wall and separates the
stellar wind from the infalling molecular cloud material. The winds in
the simulation are supersonic, and their collision against the cavity
wall generates a reverse shock close to the star ($<
500\;$AU). Because the ram pressure of the inflow/outflow is angle
dependent and the base of the cavity is subject to instabilities the
position of the reverse shock oscillates, and its shape is often
non-spherical. This is due to small fluctuations in the shape and size
of the base of the cavity wall as inflowing material is ablated and
incorporated into the outflow, and as new inflowing material
replenishes it (see Fig.~\ref{fig:snapshots}). The shear layer between
the stellar and disk winds provides a site for the growth of $\sim
700\;$AU ($10^{16}\;$cm) amplitude Kelvin-Helmholtz instabilities on
timescales of $\sim$ a few years. By $t=1200\;$yrs an instability of
this proportion can be seen driving a clump of disk wind material into
the path of the stellar wind, which leads to mass-loading of the
latter.

\begin{figure}[h]
\centering
\includegraphics[width=14cm]{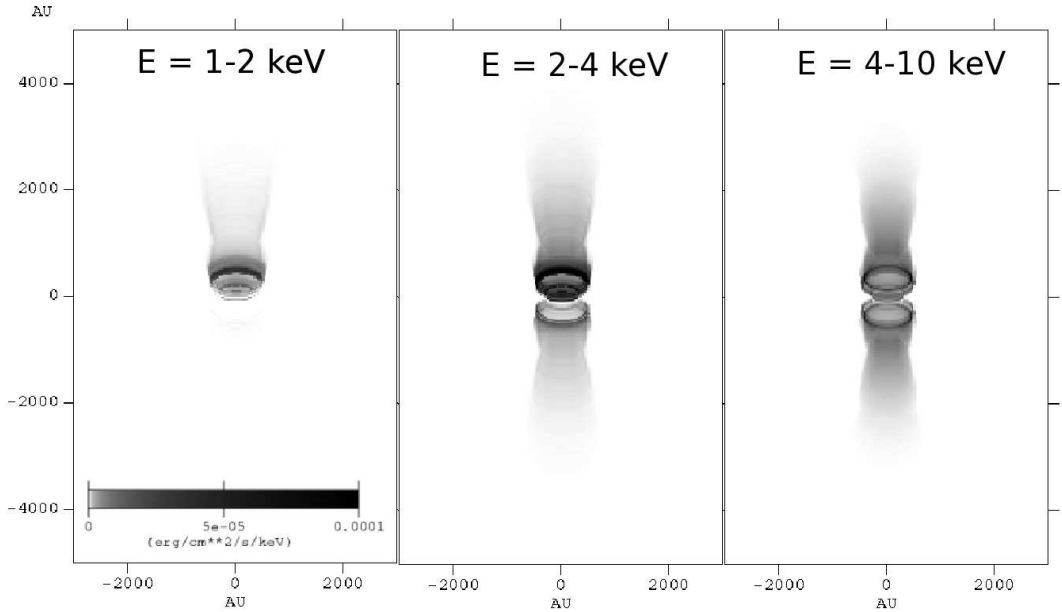}
\caption{Ray-traced synthetic broadband X-ray images.\label{fig:bbimages}}
\end{figure}

At a simulation time of $t=1000\;$yrs, the intrinsic X-ray emission
comes mainly from disturbed cloud material ($L_{\rm int_{C}}\simeq 3.3
\times 10^{33}\;{\rm erg~s^{-1}}$), then the disk wind ($L_{\rm
  int_{D}} \simeq 4.3 \times 10^{32}\;{\rm erg~s^{-1}}$), and finally
the stellar wind ($L_{\rm int_{S}}\simeq 5 \times 10^{30}\;{\rm
  erg~s^{-1}}$). The shock driven into the cloud is too slow to heat
gas up to X-ray emitting temperatures, but large quantities of cloud
material are ablated by the outflowing disk wind and mixed into this
hotter flow\footnote{Note that some heating of the cloud material may
  occur due to numerical heat conduction between the hot postshock
  disk wind and the cooler cloud material (Parkin \& Pittard
  2010).}. This process heats the cloud material to temperatures where
(soft) X-rays are emitted. However, the difference between the
\textit{attenuated} luminosities of the cloud and stellar wind
components is very small ($L_{\rm att_{C}} \simeq 8\times10^{28}\;{\rm
  erg~s^{-1}}$ and $L_{\rm att_{S}} \simeq 7\times10^{28}\;{\rm
  erg~s^{-1}}$, respectively). The explanation is that the stellar
wind emission is harder and extends to higher energies, and is less
affected by attenuation. In contrast, the cloud material, which is
heated to lower temperatures, emits prolifically at low energies and
has a much higher intrinsic luminosity, but its emission is subject to
considerable attenuation at $E < 1\;$keV. Although some variation in
X-ray emission occurs due to fluctuations in the position of the
reverse shock, these values are indicative of the mean luminosities
from the simulation.

The total attenuated luminosity from the model is $L_{\rm att_{tot}}
\simeq 2 \times 10^{29}\;{\rm erg~s^{-1}}$ which, when placed at a
distance of 1 kpc and convolved with the {\it Chandra} effective area,
equates to a count rate of $\simeq 0.1\;{\rm ks^{-1}}$. Interestingly,
this is in approximate agreement with X-ray count rates of $0.166\pm
0.041\;{\rm ks^{-1}}$ and $0.30\pm0.11\;{\rm ks^{-1}}$ inferred from
observations of Mon R2 IRS 3A by Preibisch et al. (2002) and for S106
IRS4 by Giardino, Favata, \& Micela (2004), respectively.

Examining synthetic broadband X-ray images calculated from the
simulation output one can see that the observable X-ray emission in
the 1-2, 2-4, and 4-10 keV bands originates from similar regions of
the cavity (Fig.~\ref{fig:bbimages}). The peak in the intensity in the
three bands have common origins near the reverse shock, and are mainly
generated by shocked stellar and disk wind material. Importantly, the
spatial extents of the detectable emission in all three energy bands
(1-2, 2-4, and 4-10 keV) are just below the resolution limit of
\textit{Chandra} ($\sim0.5''$) and so this model is consistent with
the unresolved nature of real MYSOs.

\section{Conclusions}
\label{sec:conclusions}

The wind-cavity interaction around an embedded MYSO has been studied
using hydrodynamic simulations and X-ray calculations. In summary, the
collision of the winds against the cavity wall generates a reverse
shock close to the star ($< 500\;$AU). The shock heated gas produces
X-ray emission with an integrated count rate ($\simeq 0.1\;{\rm
  ks^{-1}}$) and spatial extent ($< 0.5''$) in agreement with
observations of MYSOs by \textit{Chandra}.

We close with a note that future X-ray satellites, such as the
International X-ray Observatory (IXO), may have the potential to
resolve the spatial extent of the X-ray emission from the winds-cavity
interaction. If this were the case, the winds-cavity model could be
used in concert with detailed analysis at infra-red wavelengths to
place unsurpassed constraints on the parameters of MYSO outflows.

%
%
\section*{Acknowledgements}
ERP thanks The University of Leeds and PRODEX for funding. 
%
%
\footnotesize
\beginrefer
\refer {Alvarez} C., {Hoare} M., \& {Lucas} P. 2004, A\&A, 419, 203

\refer {Banerjee} R. \& {Pudritz} R.~E. 2006, ApJ, 641, 949

\refer {Banerjee} R. \& {Pudritz} R.~E. 2007, ApJ, 660, 479

\refer {Blondin} J.~M., et~al. 1990, ApJ, 356, 591

\refer Broos P.~S., et al. 2007, ApJS, 169, 353 

\refer {Bunn} J.~C., {Hoare} M.~G., \& {Drew} J.~E. 1995, MNRAS, 272, 346

\refer {Colella} P. \& {Woodward} P.~R. 1984, J.~Comput.~Phys., 54, 174

\refer {Drew} J.~E., {Bunn} J.~C., \& {Hoare} M.~G. 1993, MNRAS, 265, 12

\refer {Drew} J.~E., {Proga} D., \& {Stone} J.~M. 1998, MNRAS, 296, L6

\refer {Garay} G. \& {Lizano} S. 1999, PASP, 111, 1049

\refer {Giardino} G., {Favata} F., \& {Micela} G. 2004, A\&A, 424, 965

\refer {Hoare} M.~G. 2006, ApJ, 649, 856

\refer {Hoare} M.~G., {Drew} J.~E., {Muxlow} T.~B., \& {Davis} R.~J. 1994, ApJ, 421, L51

\refer {Liedahl} D.~A., {Osterheld} A.~L., \& {Goldstein} W.~H. 1995, ApJL, 438,
  L115

\refer {Owocki} S.~P., {Castor} J.~I., \& {Rybicki} G.~B. 1988, ApJ, 335, 914

\refer {Parkin} E.~R., \& {Pittard} J.~M. 2010, MNRAS, 406, 2373

\refer {Parkin} E.~R., {Pittard} J.~M., {Hoare} M.~G., {Wright} N.~J., \& {Drake} J.~J., 2009, MNRAS, 400, 629

\refer {Preibisch} T., {Balega} Y.~Y., {Schertl} D., \& {Weigelt} G. 2002, A\&A, 392, 945

\refer {Proga} D., {Stone} J.~M., \& {Drew} J.~E. 1998, MNRAS, 295, 595

\refer {Reipurth} B. \& {Bally} J. 2001, ARA\&A, 39, 403

\refer {Sim} S.~A., {Drew} J.~E., \& {Long} K.~S. 2005, MNRAS, 363, 615

\refer {Terebey} S., {Shu} F.~H., \& {Cassen} P. 1984, ApJ, 286, 529

\refer {Wang} J., {Townsley} L.~K., {Feigelson} E.~D., {Getman} K.~V., {Broos}
  P.~S., {Garmire} G.~P., \& {Tsujimoto} M. 2007, ApJS, 168, 100

\endrefer           
\end{document}